\documentclass{article}
\usepackage{natbib, epsfig, emulateapj, apjfonts}
\def\be{\begin{equation}}
\def\ee{\end{equation}}

\newbox\grsign \setbox\grsign=\hbox{$>$} \newdimen\grdimen \grdimen=\ht\grsign
\newbox\simlessbox \newbox\simgreatbox \newbox\simpropbox
\setbox\simgreatbox=\hbox{\raise.5ex\hbox{$>$}\llap
     {\lower.5ex\hbox{$\sim$}}}\ht1=\grdimen\dp1=0pt
\setbox\simlessbox=\hbox{\raise.5ex\hbox{$<$}\llap
     {\lower.5ex\hbox{$\sim$}}}\ht2=\grdimen\dp2=0pt
\setbox\simpropbox=\hbox{\raise.5ex\hbox{$\propto$}\llap
     {\lower.5ex\hbox{$\sim$}}}\ht2=\grdimen\dp2=0pt

\begin{document}

\title{
Ejection of high-velocity stars from the Galactic Center by an inspiralling 
Intermediate-Mass Black Hole.
}

\author{Yuri Levin $^{1,2}$}

\affil{$^1$Canadian Institute for Theoretical Astrophysics,
60 St. George Street, Toronto, ON M5S 3H8, Canada, and
\newline $^2$Leiden Observatory, P.O.Box 9513, NL-2300 RA, Leiden, The Netherlands}

\begin{abstract}
The presence  of   young stars in the immediate vicinity and strong
tidal field of  SgrA*  
remains unexplained. One currently popular idea for their origin posits
that the stars were bused in by
an Intermediate-Mass Black Hole (IMBH) which has inspiraled into
the Galactic Center a few million years ago.
 Yu and Tremaine (2003) have argued that
in this case  some of the old stars in the SgrA* cusp
would be ejected by hard gravitational collisions with the IMBH.
Here we derive a
general expression for the phase-space distribution of the ejected high-velocity
stars, given the distribution function of the stars in the cusp.
We compute it explicitly for the Peebles-Young distribution function of the 
cusp, and make a detailed model for the time-dependent ejection of stars during
the IMBH inspiral. We find that (1) the stars are ejected in a burst lasting
a few dynamical friction timescales; if the ejected stars are detected by 
Gaia they are likely to be produced by a single inspiral event, (2) if the
inspiral is circular than in the beginning
of the burst the velocity vectors of the ejected stars cluster
around the inspiral plane, but rapidly isotropise as the burst proceeds,
(3) if the inspiral is eccentric, then the stars are ejected in a broad jet
roughly perpendicular to the Runge-Lenz vector of the IMBH orbit. In a typical
cusp the orbit will precess with a period of $\sim 10^5$ years, and the
rate of ejection into our part of the Galaxy (as defined by e.g.~the Gaia
visibility domain) will be modulated periodically.
Gaia, together with the ground-based follow-up observations, will be able to
clock many high-velocity stars back to their ejection from the Galactic Center,
thus measuring some of the above phenomena. This would provide a clear signature
of the IMBH inspiral in the past 10---20 Myr.

\end{abstract}



\section{Introduction}
The origin of young stars near SgrA* has been  a major puzzle of the Galactic Center (GC)
for over a decade (Phinney 1989, Sanders 1992, Morris 1993). One of the few promising
explanations assumes that the stars were originally bound to an Intermediate-Mass
 Black Hole (IMBH) which has inspiraled from large distances (few parsecs) to the immediate
vicinity ($\sim 0.01$pc) of SgrA* (Hansen and Milosavljevic, 2003).
In this scenario the IMBH itself would be a compact remnant of a superstar of
several thousand solar masses. This superstar could be a product of stellar mergers at the center
of dense young massive cluster (Portegies Zwart et.~al.~2004, 
Gurkan et.~al.~2004) or could be
produced in a fragmenting massive accretion disc (Levin and Beloborodov 2003, Levin 2003, Goodman and Tan 2004).

Regardless of its origin, the inpiraling IMBH will eject a number of stars from the GC
and send them flying through (and  out of) the Galaxy at high velocities (Yu and Tremaine 2003, see also Gualandris et.~al.~2005).
 Brown et.~al.~2005  have recently
identified a high-velocity star in SDSS data 
and have argued that it may have been ejected from the GC. The European satellite Gaia,
scheduled to
fly in 2012, will detect and measure proper motions of  stars
to magnitude 20; the Sun has
magnitude of 19.6 at $10$kpc from Earth (Perryman et.~al.~2001). 
The threshold for radial velocity
measurement by Gaia is higher--magnitude 17; however the 
few high-proper-motion stars detected by Gaia could be observed
by the ground-based telescopes and their radial velocities could
be determined.
Thus, many high proper-motion-velocity stars could potentially be detected, their spatial position in the
Galaxy could be measured and their radial velocities could be determined from follow-up observations.
With their 6-d phase-space information determined, their trajectories could be traced back to the GC
and their times of ejection could be approximately determined;
this point was recently and independently emphasized by
Gnedin et.~al.~(2005). In this paper we derive the phase-space
distribution of the ejected stars, and explain qualitatively how 6-d information could be used to constrain 
the IMBH orbital evolution. In the next paper we will carry out the rigorous modeling of the signal expected
by GAIA from an inspiraling IMBH.

\section{Distribution of ejected stars}
The typical mass ratio of an IMBH and SgrA* ranges from $10^{-2}$ to $10^{-4}$. Therefore
the gravitational interaction of a star with the IMBH which results in the star's ejection 
occurs very close to the IMBH and can be considered instantaneous, in a sense that the IMBH
velocity is not affected by the central black hole (SgrA*) during the interaction. Moreover, the
IMBH has an immediate effect on the distribution function of stars in the cusp only in it's immediate
vicinity; therefore  the stars interacting with the IMBH have initial velocities drawn from
the distribution function of the cusp at the instantaneous location of the IMBH.  
The differential cross-section
of instantaneous encounters and the associated rates of stellar ejections were derived by Michel Henon
in a series of seminal papers (1960a,b, 1969, see also Lin and Tremaine 1980). 
Our calculations are different from Henon's, since we care about the distribution of the velocity vectors of ejected
stars, and not just their energies.
Also, we consider a single ejector (the IMBH) and do not average over the population of ejectors. The 
derivations below are of geometric nature; they are informed by but do not follow Henon's work.

Consider an IMBH flying through the stellar cusp with velocity $\vec{v}_{\rm bh}$ at radius $r$ from SgrA*.
It is reasonable to assume that the cusp is composed of stars bound to the central black hole. Therefore,
in velocity space, all the stars interacting with 
the IMBH are inside the sphere bounded by the escape velocity at $r$,
\begin{equation}
v_{\rm esc}=\sqrt{2GM\over r};
\end{equation}
see Fig.~1. In the equation above  we have assumed that the potential is dominated by central black hole
of mass $M$, which is  an excellent approximation at the location of the
puzzling young stars.

First, let us calculate the  rate of ejection of stars with velocity $v>v_{\rm ej}$, where
\begin{equation}
v_{\rm ej}=q\times v_{\rm esc},
\label{q}
\end{equation}
and $q>1$. Here and further $v$, $v_{\rm esc}$, etc.~stand for the magnitudes of the corresponding velocity vectors.
Figure 1 illustrates the velocity-space of the problem at hand. Let $\vec{v}_{\rm in}$ and $\vec{v}_{\rm out}$ be the velocity
vector of a star before and after scattering by the IMBH, respectively. It is convenient to consider
the problem in the frame of reference of the IMBH, and define
\begin{eqnarray}
\vec{w}_{\rm in}&=&\vec{v}_{\rm in}-\vec{v}_{\rm bh},\nonumber\\
\vec{w}_{\rm out}&=&\vec{v}_{\rm out}-\vec{v}_{\rm bh},\label{inout}\\
w&=&\left|\vec{w}_{\rm in}\right|=\left|\vec{w}_{\rm out}\right|.\label{w}
\end{eqnarray}

The dashed region of the velocity space marks the
stars which can be scattered to velocities $v_{\rm out}>v_{\rm ej}$; analytically it is defined by two conditions:
\begin{eqnarray}
v_{\rm in}&<&v_{\rm esc},\nonumber \\
w&>&v_{\rm ej}-v_{\rm bh}\label{ineq1}.
\end{eqnarray}
The vector of the outgoing relative velocity $\vec{w}_{\rm out}$ must satisfy
\begin{equation}
\left|\vec{v}_{\rm bh}+\vec{w}_{\rm out}\right|>v_{\rm ej},
\end{equation}
which can be translated into a 
constraint on the angle $\phi_{\rm out}$ between $\vec{v}_{\rm bh}$
and $\vec{w}_{\rm out}$:
\begin{equation}
\cos\phi_{\rm out}>\cos\phi_0={v_{\rm ej}^2-v_{\rm bh}^2-w^2\over 2v_{\rm bh}w}.
\label{phiout}
\end{equation}
The rate of ejection with velocities greater than $v_{\rm ej}$ is  given by
\begin{equation}
R(v_{\rm out}>v_{\rm ej})=\int d^3 \vec{v}_{\rm in}f(\vec{v}_{\rm in})wA(\vec{w}_{\rm in},\phi_0),
\label{rate1}
\end{equation}
where the integral is performed over the dashed volume in Fig.~1, $f(\vec{v})$ is the velocity distribution function,
and $A(\vec{w}_{\rm in}, \phi_0)$ is the cross-section for the star with initial relative velocity $\vec{w}_{\rm in}$ to be 
scattered into the velocity cone described by Eq.~(\ref{phiout}).  This 
cross-section is evaluated in the Appendix:
\begin{equation}
A(\vec{w}_{\rm in}, \phi_0)=\pi(Gm/w^2)^2{\sin^2\phi_0\over(\cos\phi_{\rm in}-\cos\phi_0)^2},
\label{A}
\end{equation}
where $m$ is the mass of the IMBH and $\phi_{\rm in}$ is the angle between $\vec{v}_{\rm bh}$ and $\vec{w}_{\rm in}$.

Now, let us find the differential ejection rate $dR(\vec{v}_{\rm out})/dV$, where $dV$ is the incremental volume
of the velocity space centered on $\vec{v}_{\rm out}$ and $dR$ is the rate of ejections of stars into this volume.
It is convenient to express $dV$ as $w^2\times dw\times d\Omega_{\rm out}$ where 
$w=|\vec{w}_{\rm out}|=|\vec{v}_{\rm out}-\vec{v}_{\rm bh}|$
and $d\Omega_{\rm out}$ is the increment of the solid angle around the vector  
$\vec{w}_{\rm out}$. The differential ejection rate is then given by
\begin{equation}
dR(\vec{v}_{\rm out})/dV=w\int f(\vec{v}_{\rm bh}+\vec{w}_{\rm in})d\Omega_{\rm in}{d\sigma (\vec{w}_{\rm in},
\vec{w}_{\rm out})\over d\Omega_{\rm out}}.
\label{diffrate}
\end{equation}
Here $d\sigma (\vec{w}_{\rm in},\vec{w}_{\rm out})/d\Omega_{\rm out}$ is the differential crossection
for gravitational scattering from $\vec{w}_{\rm in}$ to $\vec{w}_{\rm out}$, and $d\Omega_{\rm in}$ is the solid
angle increment of the integrated vector variable $\vec{w}_{\rm in}$.
The differential ejection rate is particularly easy to evaluate if the distribution function $f(\vec{v}_{\rm in})$
is axially symmetric about the vector $\vec{v}_{\rm bh}$. The ejection rate is then also axially symmetric. By averaging
Eq.~(\ref{diffrate}) over
velocity vectors $\vec{v}_{\rm out}$ which are connected by rotation about $\vec{v}_{\rm bh}$ one can show
that
\begin{equation}
dR(\vec{v}_{\rm out})/dV=w\int_{\phi_1}^\pi f(w,\phi_{\rm in})\sin \phi_{\rm in} {\partial A(w,\phi_{\rm in},\phi_{\rm out})
\over \partial \cos \phi_{\rm out}} d\phi_{\rm in},
\label{diffrate1}
\end{equation}
where
\begin{equation}
\cos\phi_1={v_{\rm esc}^2-v_{\rm bh}^2-w^2\over 2wv_{\rm bh}}.
\label{phi1}
\end{equation}
Here $f(w,\phi_{\rm in})\equiv f(\vec{v}_{\rm bh}+\vec{w}_{\rm in})$, and $A(w,\phi_{\rm in},\phi_{\rm out})\equiv A(\vec{w}_{\rm in},
\phi_{\rm out})$ as given by Eq.~(\ref{A}).

For a  general distribution function $f(\vec{v})$, the integrals in Eqs.~(\ref{rate1}) and (\ref{diffrate1}) must be integrated numerically.
In the next section, we consider an important particular case where these integrals can be  evaluated analytically.

\section{IMBH inspiraling through  the $\rho\propto r^{-1.5}$ cusp}
A commonly considered cusp distribution 
function which is convenient for analytical work  is the Peebles-Young
distribution function $f(v)=f_0$; this corresponds to an isotropic cusp with the density profile $\rho\propto r^{-1.5}$
(Young 1980).
This profile
is also very close to that of the observed cusp in the Galactic Center (Genzel et.~al.~2003; see below), and
hence in this section we will consider in some detail the distribution of the stars ejected by the  IMBH inspiraling
through the Peebles-Young cusp.
 
The ejection rate is obtained by evaluating the 
integral in Eq.~(\ref{rate1}). We have found $w$ and $\phi_{\rm in}$ to be
 convenient integration variables; it is easiest to integrate first over $\phi_{\rm in}$ and then
over $w$.  One then finds the following for the ejection rate:
\begin{equation}
R(v_{\rm out}>v_{\rm ej})={3\pi\over 2}{n(Gm)^2\over v_{\rm esc}^3}\Lambda(p,q),
\label{rate2}
\end{equation}
where $p=v_{\rm bh}/v_{\rm esc}<1$, $q=v_{\rm ej}/v_{\rm esc}>1$, $n$ is the local number density of stars and
\begin{eqnarray}
\Lambda(p.q)&=&{(1+2p-q)\over p}\left[{p^2+q^2\over q^2-1}-{1\over 2}\right]\\
            &+&\left[{1\over q-p}-{1\over 1+p}\right]{q^2-p^2\over 2p}\left[1-{q^2-p^2\over q^2-1}\right]\nonumber\\
            &-&{(1+p)^3-(q-p)^3\over 6p(q^2-1)}-\ln \left({1+p\over q-p}\right).\nonumber\label{lambda}
\end{eqnarray}
The differential ejection rate is then obtained by evaluating the integral in Eq.~(\ref{diffrate1}), and is given by
\begin{eqnarray}
dR(\vec{v}_{\rm out})/dV&=&{\pi f (Gm)^2\over w^3}\left\{2C\left({1\over C-C_1}-{1\over C+1}\right)\right.\nonumber\\
                        &+&\left.{1-C^2\over 2}\left[{1\over (C-C_1)^2}-{1\over (C+1)^2}\right]\right\},\label{diffrate2}
\end{eqnarray}
where
\begin{eqnarray}
w&=&\sqrt{v_{\rm out}^2+v_{\rm bh}^2-2v_{\rm out}v_{\rm bh}\cos\theta},\nonumber\\
C&=&(v_{\rm out}\cos\theta-v_{\rm bh})/w,\\
C_1&=&{v_{\rm esc}^2-v_{\rm bh}^2-w^2\over 2wv_{\rm bh}},\nonumber
\end{eqnarray}
and $\theta$ is the angle between $\vec{v}_{\rm out}$ and $\vec{v}_{\rm bh}$.

Let us put make a numerical estimate for Eq.~(\ref{rate2}) with the parameters relevant for the
Galactic Center.  We are interested in stars with
luminosity no less than solar, since we want the star to be observable by Gaia at the distance of $\sim 10$ kpc.
 At the moment the observational constraint on the number of such stars in the central cusp
is poor, as the faintest stars observable by Genzel's and Ghez' groups with the VLT
and Keck have masses $>5M_{\odot}$ (R.~Genzel and A.~Ghez  2005, private communications).
 Among the stars which are observable, the mass function does not follow Salpeter's law,
and is clearly top-heavy. The IMF favoring high masses
is consistent both with stars born in a dense  accretion disc (Artymowicz et.~al.~1993, Levin 2003, Levin and Beloborodov 2003) or
brought in as a part of a core of a dense young cluster (Gerhard 2001). To make further progress we need to
parametrize the distribution of stars with masses $\sim M_{\odot}$ in the central cusp:
\begin{equation}
n(r)=8\times 10^6\hbox{pc}^{-3}n_0(0.1\hbox{pc}/r)^{1.5+\delta}.
\label{cusp}
\end{equation}
The slope $\delta$ was measured by Genzel et.~al.~(2003) and Shodel et.~al.~(2005) to be $-0.2$ and $0.2$, respectively,
with both of the analyses assuming constant mass-to-light ratio. Thus the cusp slope is close to the Peebles-Young profile,
and in the fiducial example below we will assume $\delta=0$. As explained above, the normalization factor $n_0$ is not
well known at present;
$n_0=1$ corresponds to the case when the mass density at $0.1$pc quoted in Eq.~(2)
of Genzel et.~al.~(2003) 
is dominated by solar-mass stars. With this parametrization the ejection rate becomes
\begin{equation}
R(v_{\rm out}>v_{\rm ej})\simeq 60\hbox{Myr}^{-1}n_0\times \left({m/M\over 0.001}\right)^2\times
              \left({M\over 4\times 10^6M_{\odot}}\right)^{1/2}\Lambda(p,q).
\label{rate2.5}
\end{equation}

Consider now an IMBH of mass $m=10^{-3}M=4000M_{\odot}$ which is inspiraling into SgrA* on a circular orbit.
For a circular IMBH orbit, $p=1/\sqrt{2}$ and the function $\Lambda(1/\sqrt{2},q)$ is plotted as a function
of $q$ in Fig.~2. For the Peebles-Young profile the inspiral proceeds at constant rate, and
the radius and the escape velocity of the orbit are given by
\begin{eqnarray}
r&=&r_0\exp(-t/t_{\rm friction}),
\label{rinsp}\\
v_{\rm esc}&=&\sqrt{2GM/r_0}\exp(t/2t_{\rm friction})\nonumber
\end{eqnarray} 
where
\begin{equation}
t_{\rm friction}={M^{1.5}\over \sqrt{G}m\rho r^{1.5}Q}\simeq 10^6\hbox{yr}\times (5/n_0Q)
\label{tfriction}
\end{equation}
is the dynamical friction timescale, with $Q$ being the Coulomb logarithm [cf. Eqs.~(13) and (14)
of Levin, Wu, and Thommes 2005; our $Q$ is their $\log\Lambda$].
Let us now calculate, as a function of time, the number of  stars which are ejected during
the inspiral and which will have velocities greater than $v_{\rm galaxy}=800$km/sec
once they climb through the bulge potential.
We assume that the  escape velocity from the bottom
of the bulge is $v_{\rm bulge}=300$km/sec (see Yu and Tremaine 2003 for discussion). Therefore,
after collision with the IMBH the star must have velocity greater than
\begin{equation}
v_{\rm ej}=\sqrt{v_{\rm esc}^2+v_{\rm bulge}^2+v_{\rm galaxy}^2}.
\label{vej4}
\end{equation}
 To evaluate the number of ejected solar-type stars as a function of time passed since
the beginning of the inspiral,
we integrate Eqs.~(\ref{rate2.5}), (\ref{rinsp}), and (\ref{vej4}) with respect to the time
variable.
The result is shown in Fig.~3, for a particular case of $n_0=1$ and $Q=5$. The rate of ejections as a function
of time is shown in Fig.~4. 
The solid part of the curve, i.e.~the part with $t<3.25$Myr, marks the first stage  of the inspiral when
the IMBH mass is smaller than the mass of the stellar cusp out to the radius of the IMBH orbit.
Therefore, the Black-Hole inspiral does not significantly affect the stellar distribution function
of the cusp, and our analysis is reliable. For $t>3.25$Myr, which corresponds to orbital radius $<0.015$pc,
the IMBH dominates the mass budget inside its orbit and depletes its neighborhood of stars on the timescale of
$t_{\rm friction}$. Therefore, within our analytical formalism we cannot reliably predict the rate of ejections since
the original distribution function of the cusp is strongly altered by the IMBH\footnote{Yu and Tremaine (2003) have estimated
the rate of ejections in this regime by appealing to  the loss-cone formalism. They have assumed that the BH binary interacts with the
stars on plunging orbits and that the star with pericenter radius similar to that of the binary gets ejected almost immediately.
The validity of both assumptions is unclear for the small ($10^{-3}$) mass ratios considered here.} . We   model very  roughly
this IMBH-dominated regime by positing  that the overall density of stars and the rate of inspiral are
 declining exponentially on the timescale
of $t_{\rm friction}$, while assuming that the overall form of the cusp distribution function remains the same.
The dashed part of the curve shows the number of ejections calculated within this rough model.
To get a better intuition for the parameters of ejected stars, we have generated mock data based on the ejection rates computed above.
We present this data as a scatter plot in Fig.~5: on the $x$-axis we show the time of ejection and on the 
$y$-axis we show the velocity of the ejected star after it escapes from the bulge.
With Gaia one will be able to clock the high-velocity stars back
to their ejections  and thus  measure both $t_{\rm ejection}$ and 3-d velocity for many of the
high-velocity  stars within 10 kpc from the Earth whose proper-motion vector is pointing to the Galactic Center.
Building a scatter plot like the one in Fig.~5 for the ejected high-velocity stars will provide
a clear diagnostic for the IMBH inspiral as the origin of the ejected stars.

On the theoretical side, we have a problem-more than half of the ejected stars are produced during the part of the inspiral which
cannot be modeled analytically. Therefore, we have conceived, but not yet carried out, a series of numerical experiments aimed at addressing the
part of the inspiral where IMBH mass dominates that of the stellar cusp. For now, we limit ourselves to two qualitative remarks:\newline
(a) The number of the stars ejected during a single inspiral
is dependent on the poorly-known distribution of stars in the cusp but may be counted in hundreds.
These stars are potential targets for GAIA.\newline
(b) The stars are ejected in a burst the most intense part of which lasts 2---3 dynamical friction
timescales. If no more than one IMBH per 10 Myr inspirals into SgrA*, than the high-velocity stars 
seen by GAIA should be all generated by a single inspiral event.

How anisotropic are the ejected stars? First we must note that the velocity vector $\vec{v}_{\rm star}$ of a star after
it escapes the potential of the central black hole is  different from $\vec{v}_{\rm out}$, which is the stellar velocity right
after gravitational collision with the IMBH. Generally the velocity of the escaped star
\begin{equation}
\vec{v}_{\rm star}=\vec{v}_{\rm star}(\vec{r},\vec{v}_{\rm out})
\label{vstares}
\end{equation}
can be found by solving for the Keplerian hyperbolic trajectory of the star; here $\vec{r}$ is the position
of the IMBH during the gravitational collision. The mathematical expressions for Keplerian hyperbolae can be
found in a standard textbook on Celestial Mechanics, e.g.~Danby (1988).

A useful way to quantify the anisotropy of ejections is to compute the dispersion of angles $\chi(\vec{v}_{\rm star})$
between the velocity vectors of the ejected stars and the inspiral plane. This dispersion $\sigma$ can be calculated by evaluating the
following integral over a single period of the IMBH orbit:
\begin{equation}
\sigma^2={1\over N_{\rm ej}(T)}\int_0^T dt\int d^3\vec{v}_{\rm out}\times {dR(\vec{v}_{\rm bh}(t),\vec{v}_{\rm out})\over dV}
\times \chi^2[\vec{v}_{\rm star}(\vec{r}(t),\vec{v}_{\rm out})],
\label{sigma}
\end{equation}
where $t$ is the time, $T$ is the period of the orbit, $dR/dV$ is given by Eq.~(\ref{diffrate}) [and in appropriate cases
by Eq.~(\ref{diffrate1}) or (\ref{diffrate2})], 
and $N_{\rm ej}(T)$ is the number of stars ejected during an orbital period:
\begin{equation}
N_{\rm ej}(T)=\int_0^T dt\int d^3\vec{v}_{\rm out}\times {dR(\vec{v}_{\rm bh}(t),\vec{v}_{\rm out})\over dV}.
\label{Nej}
\end{equation}
In Fig.~6 we plot $\sigma$ vs time for the circular IMBH inspiral, with the thick horizontal line
corresponding to isotropic distribution of the velocity vectors. At early times the stars ejected with
velocities $>800$km/sec will be clustering towards the inspiral plane, while at later times
the velocity vectors of the ejected stars have no strong anisotropy. Unfortunately, at early times the ejection rate
is low, and thus in the example we are considering only the first 10 or so ejections will
show strong anisotropy. GAIA will see at most about 1/4 of the ejected stars; we think therefore that detection
of spatial anisotropy for circular inspiral is unlikely.

\paragraph{Eccentric inspiral.} There is a stronger chance to detect spatial anisotropy of the ejected stars
if the IMBH inspirals on a strongly eccentric orbit. Most high-velocity stars will be ejected when IMBH passes
the pericenter. These stars will then form a broad jet  which will have the general direction of the IMBH
velocity vector at the pericenter, i.e.~perpendicular to the Runge-Lenz vector of the IMBH orbit.
The elliptical orbit of the IMBH will precess in the potential of the cusp, and the jet will rotate with
the orbital ellipse. Therefore, the rate of stars ejected into the region of the Galaxy observable by
GAIA (the GAIA sphere) will be modulated with the period of the IMBH precession. {\it It is this periodic modulation
that is potentially observable, and that would be a tell-tale of the eccentric IMBH inspiral}.

For the Peebles-Young cusp the period of precession depends only on the IMBH eccentricity
and does not depend on its semimajor axis. Thus during the first stage of the inspiral, when the
cusp is not affected by the IMBH, the period of modulation will be constant and equal to
\begin{equation}
T_{\rm precession}=1.3\left(M\over 4\times 10^6 M_{\odot}\right)n_0^{-1}\times 10^5\hbox{yr},
\label{precession}
\end{equation}
see Munyaneza et.~al.~(1999). Here we have taken the IMBH eccentricity\footnote{Previous
calculations (e.g., in the Appendices of Gould and Quillen 2003 and Levin et.~al.~2005) argue
that the eccentricity of the IMBH inspiraling through the Peebles--Young cusp remains constant with
time. However, these calculations neglect secular torques exerted on a precessing eccentric orbit; therefore
the question of eccentricity evolution of an inspiraling IMBH remains open.} 
$e=0.7$. The  Galactic Center cusp has spatial distribution similar to the
Peebles-Young profile, and therefore the period of precession will be nearly
constant during the first stage of the inspiral. During the second stage, however, when the
IMBH depletes the stellar density of the cusp, the precession period becomes longer.
Detection of this period change would clearly provide a strong probe of the cusp structure.
In Fig.~7 we show the rate of stars injected into the GAIA visibility sphere as a function of the
injection time (all computed for the fiducial cusp parameters
described above). The  star is defined to be in the GAIA visibility sphere if at the current time,
taken  to be $10$Myr, the star is within $10$kpc from the Earth. The orbit of the IMBH is taken to
lie in the Galactic plane. We clearly see in Fig.~7 both the periodicity in the injection rate
and the change in this periodicity when the IMBH begins to empty its local region of the cusp.

\section{Conclusions.}
In this paper we have derived the analytic expression for the time-dependent phase-space distribution
of stars ejected from the Galactic Center as a result of the IMBH inspiral. We have discussed
how Gaia, together with the  follow-up ground-based radial velocity measurements, 
may be used to track the history of ejections from
the Galactic Center; this point was recently and independently made by Gnedin et.~al.~2005.
We find that a single IMBH inspiral produces a burst in the rate of ejection which has a duration a few
dynamical friction timescales. In the very beginning of the burst, the velocities
of the ejected stars cluster around the inspiral plane, however this anisotropy will be hard to detect
and most of the stars will be ejected isotropically.  

If the IMBH orbit is eccentric, then the ejected stars form a broad ``jet'' roughly aligned
with IMBH velocity in the pericenter. Because of the cusp potential, the IMBH orbit will precess with
the period of $\sim 10^5$ years, and the rate of stellar ejections into the GAIA visibility zone
will be strongly periodically modulated. When the IMBH begins to modify stellar distribution in its
neighborhood (i.e., when its mass becomes comparable to the mass of stars inside its orbit),
the period of precession will increase, as shown in Fig.~7.

Therefore, the burst in the ejection rate and potential modulation of this rate during
the burst, if detected by GAIA, would be a clear signature that an IMBH has inspiraled into
the Galactic Center in the past 10-20 Myr.

\acknowledgments
We thank Tim de Zeeuw for carefully reading the manuscript and making
a number of important suggestions, and Scott Tremaine for
encouraging discussions. This research was supported by NSERC.

\section*{Appendix: evaluation of the crossection {\bf $A$}}
In this Appendix we evaluate the crossection for a star with the initial velocity
$\vec{w}_{\rm in}$ relative to the IMBH to scatter into the cone given by
Eq.~(\ref{phiout})
\begin{equation}
\cos\phi_{\rm out}>\cos\phi_0={v_{\rm ej}^2-v_{\rm bh}^2-w^2\over 2v_{\rm bh}w}.
\label{phiout1}
\end{equation}
Here $\phi_{\rm out}$ is the angle between $\vec{w}_{\rm out}$ and the
velocity vector of the IMBH.

Let $P$ be the plane which is perpendicular to $\vec{w}_{\rm in}$ and which contains
the IMBH. Let $\vec{b}$ be the impact parameter vector, i.e.~a  vector directed
from the IMBH to the projection of the star onto $P$ before the star is scattered 
by the IMBH. After the scattering is complete, the stellar velocity relative to the IMBH 
is given by  
\begin{equation}
\vec{w}_{\rm out}=\vec{w}_{\rm in}\cos\theta-\vec{b}\sin\theta(w/b),
\label{wf}
\end{equation}
where $\theta$ is the angle of deflection given by
\begin{equation}
\tan(\theta/2)={Gm\over w^2b}.
\label{theta2}
\end{equation}
 From Eq.~(\ref{wf}) we see that

\begin{equation}
\cos\phi_{\rm out}=\cos{\phi_{\rm in}}\cos{\theta}+\sin{\phi_{\rm in}}\sin{\theta}
\cos{\phi_b},
\label{cosphiout}
\end{equation}
where $\phi_b$ is the polar angle in the $\vec{b}$-plane.

Equations (\ref{cosphiout}) and (\ref{theta2}) together with Eq.~(\ref{phiout1})
define a finite domain in the $\vec{b}$-plane. It is straightforward but somewhat
tedious  to
find the boundary of this domain; this calculation is not shown here. The area of
the domain is the crossection we seek and is given by
\begin{equation}
A(\vec{w}_{\rm in}, \phi_0)=\pi(Gm/w^2)^2{\sin^2\phi_0\over(\cos\phi_{\rm in}-\cos\phi_0)^2},
\label{A1}
\end{equation}
which is the Equation (\ref{A}) in the text.

\begin{figure}
\begin{center}
\plotone{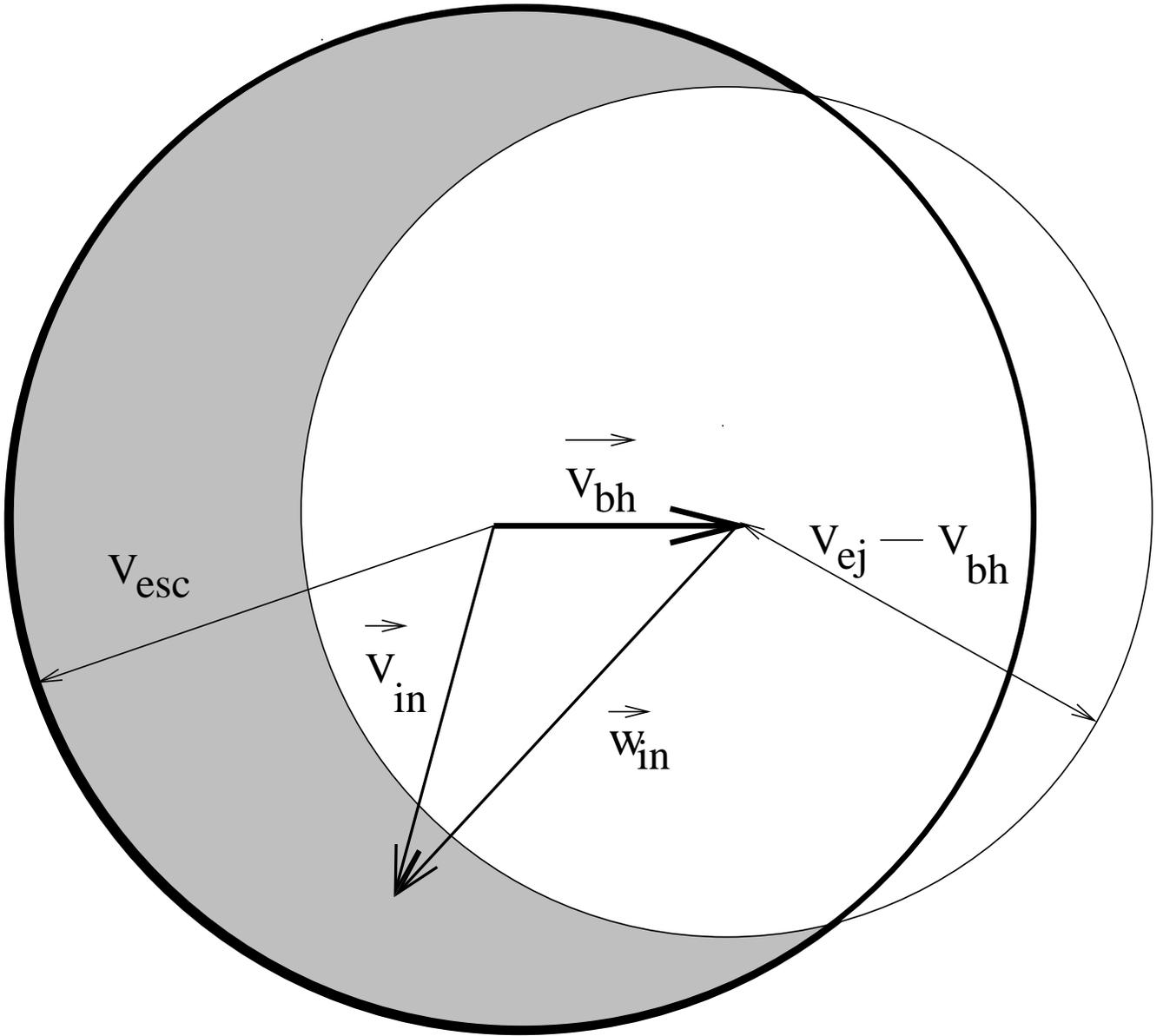}
\end{center}
\caption{Velocity-space diagram of gravitational
scattering of stars by the IMBH. The initial stellar
velocities $\vec{v}_{\rm in}$
lie inside the sphere of radius $v_{\rm esc}$.
The stars with velocities in the shadowed region of the
figure can be scattered by the IMBH to  velocities greater
than $v_{\rm ej}$. Other symbols are explained in the text.}
\end{figure}

\begin{figure}
\begin{center}
\plotone{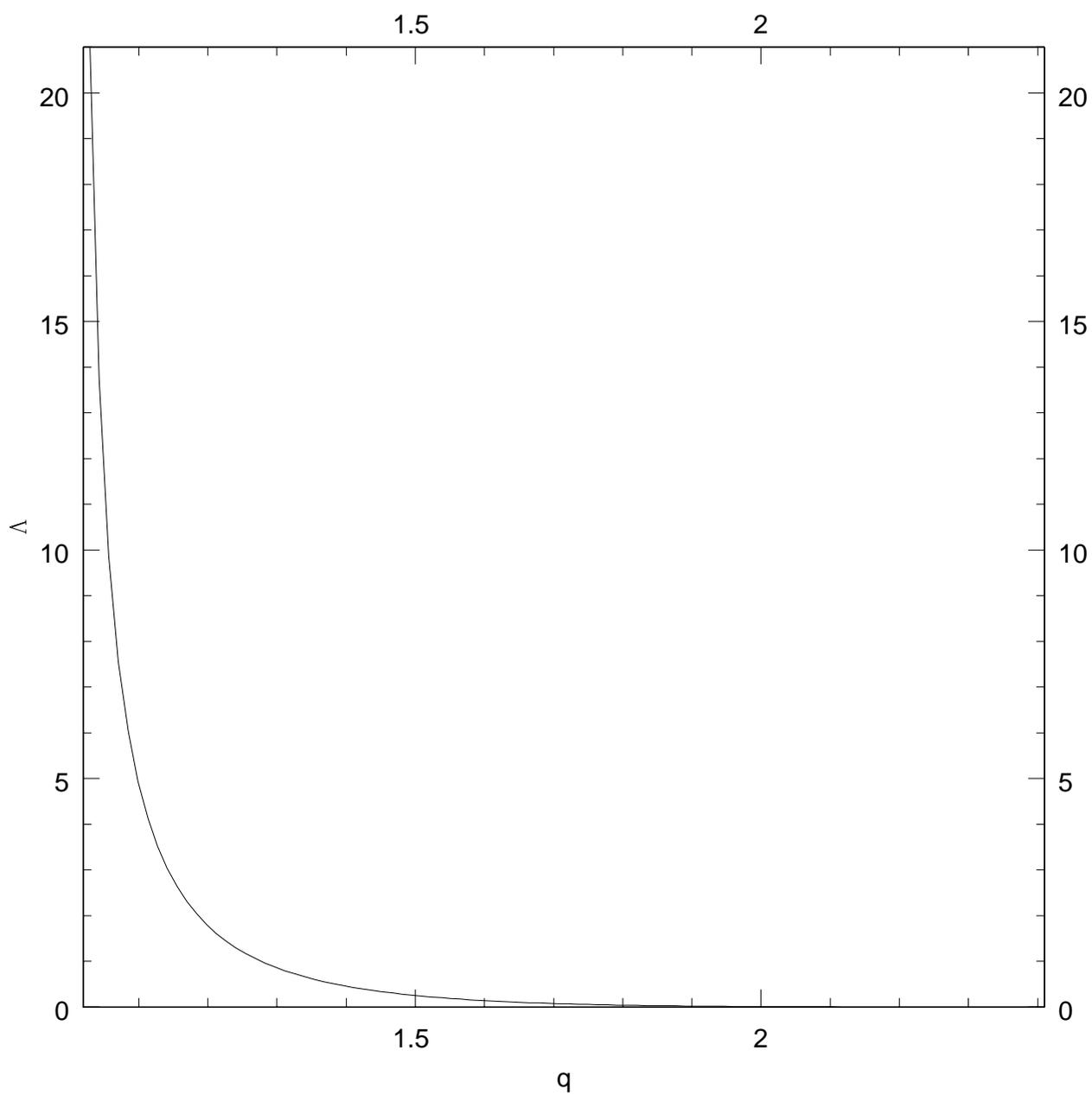}
\end{center}
\caption{The function $\Lambda(p,q)$ computed for
a circular IMBH orbit, $p=1/\sqrt{2}$}
\end{figure}

\begin{figure}
\begin{center}
\plotone{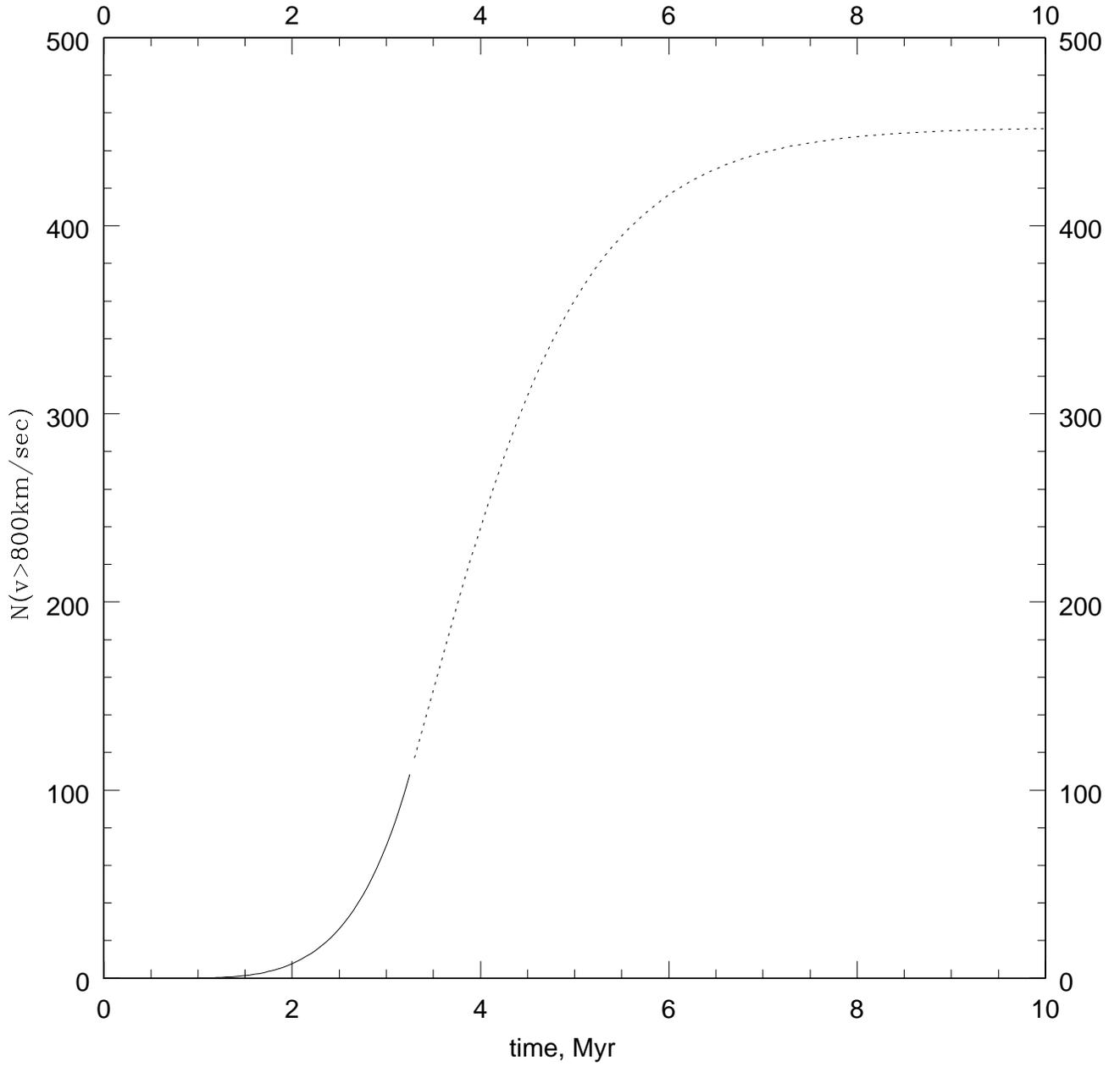}
\end{center}
\caption{The number of ejected stars as a function of
the ejection 
time. The
solid curve represents the region where analytical
calculations are appropriate, the dashed curve represents
the region where our model is a rough approximation.}
\end{figure}

\begin{figure}
\begin{center}
\plotone{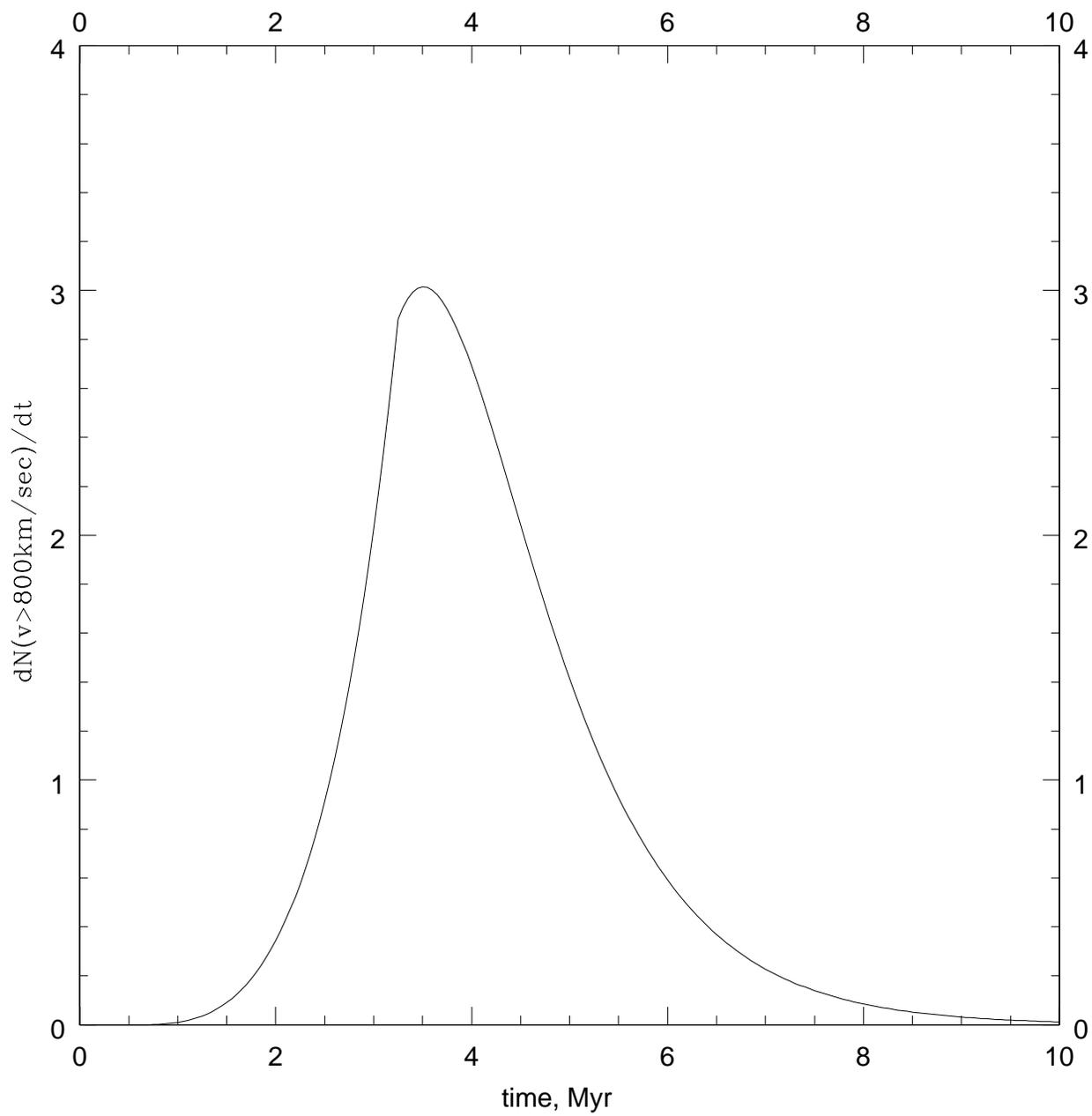}
\end{center}
\caption{The rate of ejections as a function of the ejection time.}
\end{figure}

\begin{figure}
\begin{center}
\plotone{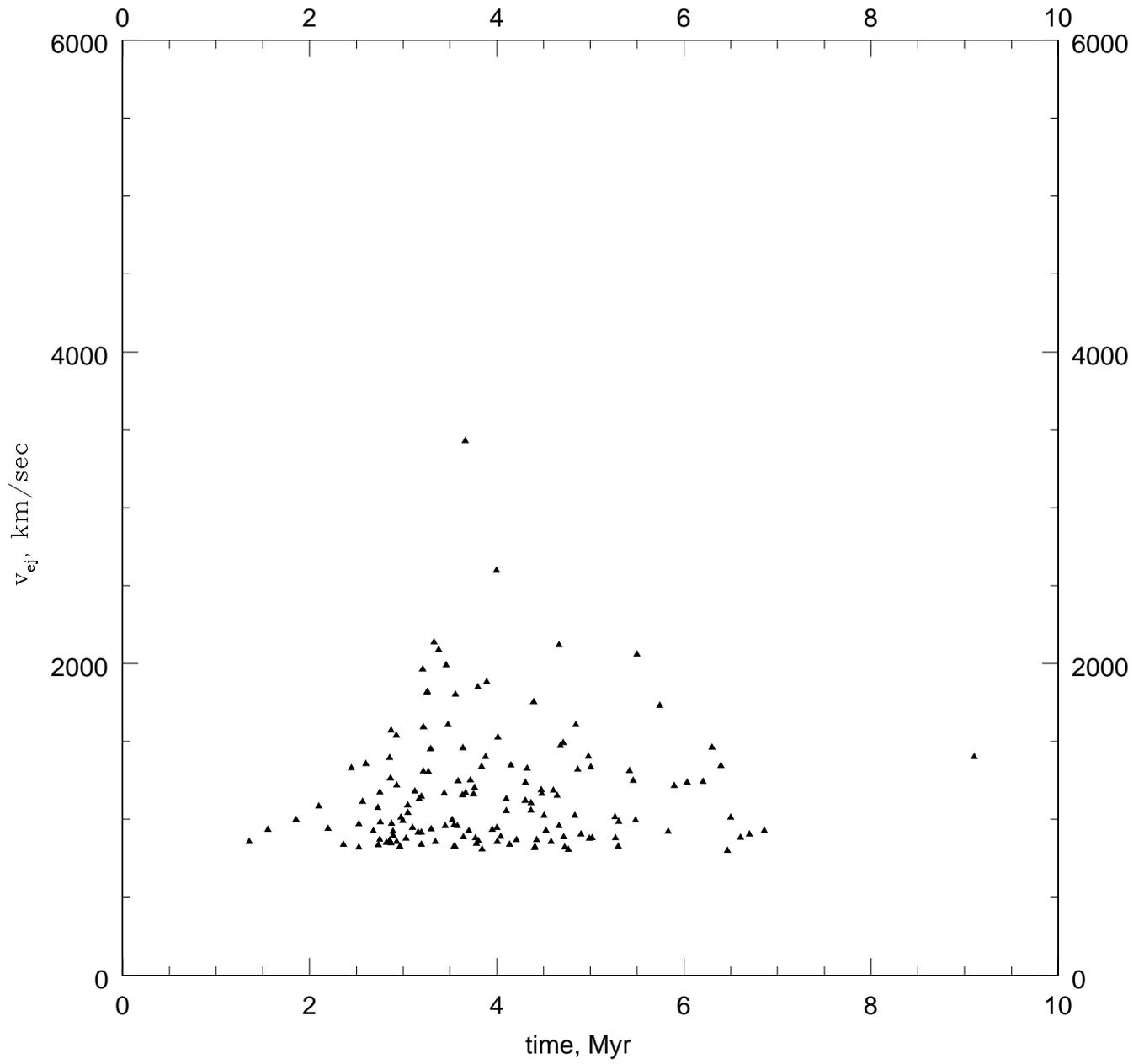}
\end{center}
\caption{Scatterplot of velocities and ejection times for
the ejected stars}
\end{figure}

\begin{figure}
\begin{center}
\plotone{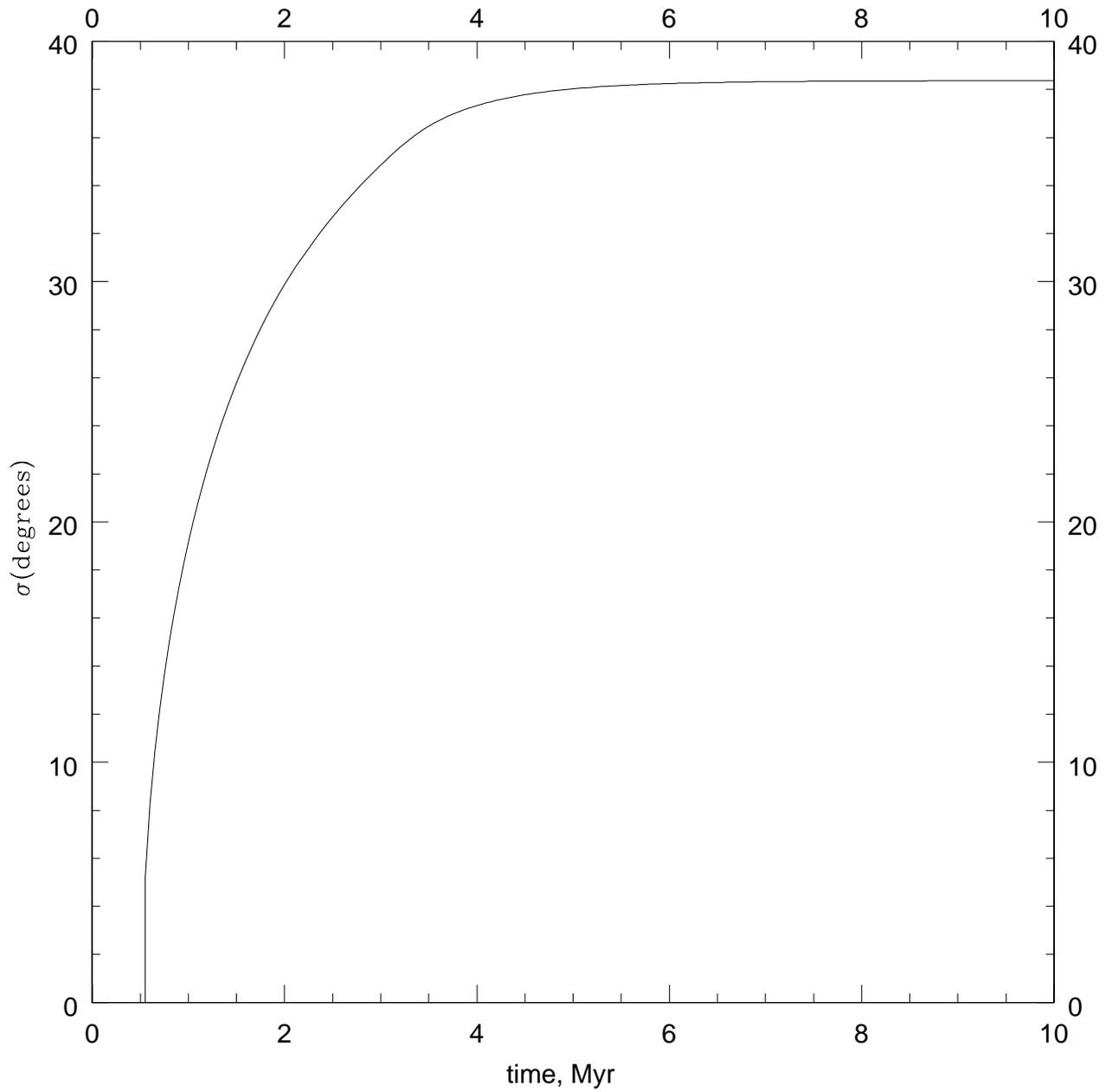}
\end{center}
\caption{Dispersion of angles between the ejection velocity vectors
and the inspiral plane plotted as a function of ejection time}
\end{figure}

\begin{figure}
\begin{center}
\plotone{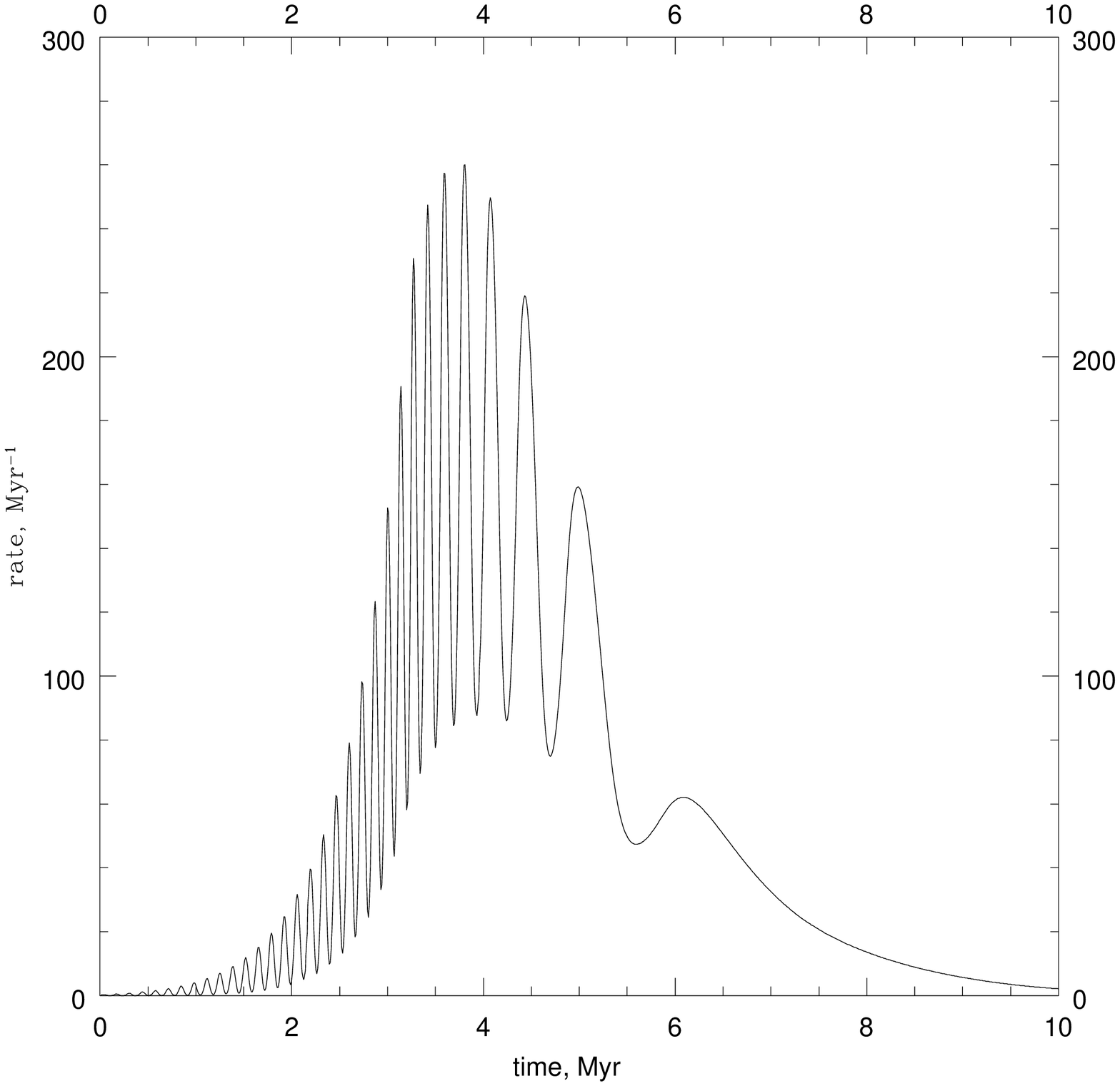}
\end{center}
\caption{Eccentric inspiral: the rate of ejections of
stars into GAIA visibility domain as a function of the ejection
time. Here the IMBH orbital eccentricity is $e=0.7$.}
\end{figure}


\begin{references}
\reference{} Artymowicz, P., Lin, D.~N.~C., \& Wampler, E.~J.~1993, ApJ, 409, 592
\reference{} Brown, W.~B., et.~al.~2005, ApJ, 622, L33
\reference{} Danby, J.~M.~A.~1988, Fundamentals of Celestial Mechanics, 
           Richmond, Va., USA, Willmann-Bell, 2nd ed.
\reference{} Gerhard, O.~2001, ApJ, 546, L39	   
\reference{} Gnedin, O.~Y., et.~al.~2005, astro-ph/0506739
\reference{} Goodman, J., \& Tan, J.~C.~2004, ApJ, 608, 108
\reference{} Gualandris, A., Portegies Zwart, S., \& Sipior, M.~2005, 
accepted to MNRAS, astro-ph/0507365
\reference{} Gurkan, M.~A., Freitag, M., \& Rasio, F.~A.~2004, ApJ, 604, 632
\reference{} Hansen, B.~M.~S., \& Milosavljevic, M.~2003, ApJ, 593, L77
\reference{} Henon, M.~1960a, AnAp, 23, 467 
\reference{} Henon, M.~1960b, AnAp, 23, 668
\reference{} Henon, M.~1969, A\&A, 2, 151
\reference{} Levin, Y.~2003, astro-ph/0307084
\reference{} Levin, Y., \& Beloborodov, A.~M.~2003, ApJ, 590, L33
\reference{} Levin, Y., Wu, A.~S.~P., Thommes, E.~W.~2005, submitted to ApJ, astro-ph/0502143
\reference{} Lin, D.~N.~C., \& Tremaine, S.~1980, ApJ, 264, 364
\reference{} Morris, M.~1993, ApJ, 408, 496
\reference{} Munyaneza, F., Tsiklauri, D., Viollier, R.~D.~1999, ApJ, 526, 744
\reference{} Phinney, E.~S.~1989, in IAU Symp.~136, The Center of the Galaxy, 
              ed.~M.~Morris (Dordrecht:Kluwer), 543
\reference{} Perryman, M.~A.~C., et.~al.~2001, A\&A, 369, 339
\reference{} Portegies Zwart, S.~F., et.~al.~2004, Nature, 428, 724
\reference{} Sanders, R.~H.~1992, Nature,  359, 131
\reference{} Shodel, R., et.~al.~2005, in preparation
\reference{} Young, P.~1980, ApJ, 242, 1232
\reference{} Yu, Q., \& Tremaine, S.~2003, ApJ, 599, 1129



\end{references}
\end{document}